# Electronic Structure Tunability by Periodic *meta*-Ligand Spacing in One-Dimensional Organic Semiconductors

Ignacio Piquero-Zulaica,[*,†] Aran Garcia-Lekue,[‡,§] Luciano Colazzo,[‡] Claudio K. Krug,[∥] Mohammed S. G. Mohammed,[‡,†] Zakaria M. Abd El-Fattah,[⊥,#] J. Michael Gottfried,[∥] Dimas G. de Oteyza,[‡,†,§] J. Enrique Ortega,[‡,†,∇] Jorge Lobo-Checa[*,○,◆]

[†]Centro de Física de Materiales CSIC/UPV-EHU-Materials Physics Center, Paseo Manuel de Lardizabal 5, E-20018 San Sebastián, Spain

[‡]Donostia International Physics Center (DIPC), Paseo Manuel de Lardizabal 4, E-20018 Donostia-San Sebastián, Spain

[§]Ikerbasque, Basque Foundation for Science, 48011 Bilbao, Spain

[∥]Fachbereich Chemie, Philipps-Universität Marburg, Hans-Meerwein-Str. 4, 35032 Marburg, Germany

[⊥]ICFO-Institut de Ciencies Fotoniques, The Barcelona Institute of Science and Technology, 08860 Castelldefels, Barcelona, Spain

[#]Physics Department, Faculty of Science, Al-Azhar University, Nasr City, E-11884 Cairo, Egypt

[∇]Dpto. Física Aplicada I, Universidad del País Vasco, E-20018 San Sebastián, Spain

[○]Instituto de Ciencia de Materiales de Aragón (ICMA), CSIC-Universidad de Zaragoza, E-50009 Zaragoza, Spain

[◆]Departamento de Física de la Materia Condensada, Universidad de Zaragoza, E-50009 Zaragoza, Spain

**ABSTRACT:** Designing molecular organic semiconductors with distinct frontier orbitals is key for the development of devices with desirable properties. Generating defined organic nanostructures with atomic precision can be accomplished by on-surface synthesis. We use this "dry" chemistry to introduce topological variations in a conjugated poly(*para*-phenylene) chain in the form of *meta*-junctions. As evidenced by STM and LEED, we produce a macroscopically ordered, monolayer thin zigzag chain film on a vicinal silver crystal. These cross-conjugated nanostructures are expected to display altered electronic properties, which are now unraveled by highly complementary experimental techniques (ARPES and STS) and theoretical calculations (DFT and EPWE). We find that *meta*-junctions dominate the weakly dispersive band structure, while the band gap is tunable by altering the linear segment's length. These periodic topology effects induce significant loss of the electronic coupling between neighboring linear segments leading to partial electron confinement in the form of weakly coupled quantum dots. Such periodic quantum interference effects determine the overall semiconducting character and functionality of the chains.

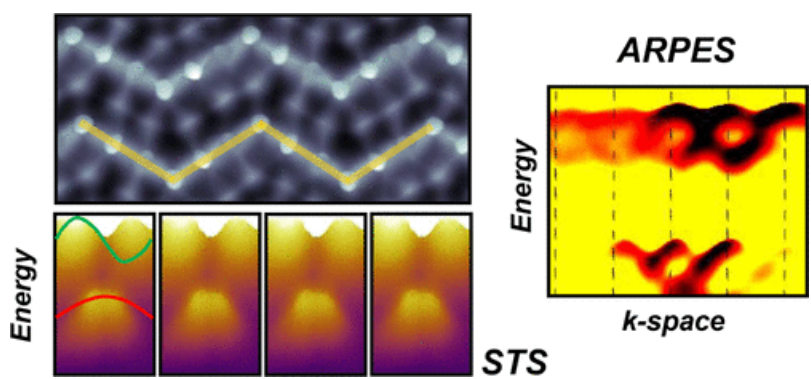

TOC

Conjugated polymers in the form of molecular chains are extensively used in industry as light emitting materials, photocatalysts, solar cells and biosensors due to their large and tunable band gaps.(1−4) Control over their electronic properties is accomplished through topological functionalization of these π-conjugated oligophenylene chains, *i.e.*, modification of their conductive pathways. In particular, changes of conjugation (from linear to cross-conjugation) by precise transitions from *para*- to *meta*-ligand substitutions(5,6) weaken the electronic communication between the repeating units of the polymer.(7) Such modifications have also been described as quantum interference electron pathways,(8−10) and bear predicted effects such as scarcely dispersive bands,(5) wider band gap,(11) distinct optical properties,(4) electronic switching capabilites(12) and low conductance properties.(8−10,13,14) However, periodic *meta*-junction zigzag chains may also show enhanced charge mobility as compared to their poly(*para*-phenylene) counterparts, reaching values comparable to those of amorphous silicon.(15) Despite this wealth of industrially attractive properties of cross-conjugated polymers, key fundamental information, such as the predicted electronic structure, awaits experimental validation. Such deficiency of fundamental knowledge limits the confidence in the existing predictions, according to which topology is expected to affect the electronic properties of the polymer. Several obstacles are responsible for the lack of the aforementioned experimental confirmation: (i) the need of generating atomically identical chains exhibiting repeated *para*- to *meta*-ligand substituted units, (ii) the synthesis of well-aligned chains, to be probed by nonlocal, averaging spectroscopies, (iii) the minimization of lateral interactions, prone to affect their intrinsic band structure, and (iv) the right choice of a support that sufficiently decouples the electronic signal from the investigated oligophenylene chains.

To overcome such obstacles, solutions can be found within the context of Surface Science. Particularly, the first prerequisite for obtaining perfectly reproducible cross-conjugated zigzag polymers can be accomplished by bottom-up on-surface synthesis. Surface-assisted C–C coupling processes have been recently applied to generate graphene nanoribbons (GNRs) with different edge terminations and widths,(16−21) and other types of oligophenylene chains.(22−25) Second, the chain alignment for nonlocal characterization can be achieved by the use of nanotemplated substrates, such as vicinal surfaces.(26−29) These special surfaces have been successfully used for the macroscopic alignment of carbon-based chains, a fundamental requirement for angle resolved photoemission (ARPES) experiments.(23,24,30,31) With respect to the minimization of lateral interchain coupling, this is an inherent feature of the Ullmann-type surface reactions(22) since the halogens are cleaved during the synthesis positioning themselves between neighboring chains.(23,24,32) These adatoms are reported to laterally decouple adjoining chains, without affecting the polymer's band structure, except for a minimal rigid energy shift similar to doping effects.(24,33) Finally, the substrate plays a fundamental role as a catalyst of the Ullmann reaction, making its choice crucial for a successful oligomer coupling. Good candidates that present excellent yields, control and reproducibility are the closed packed surfaces of coinage metals, which are extensively used for Ullmann-type surface reactions. Among these, silver stands out as a promising substrate, since it weakly interacts with the products while exhibiting large adsorbate diffusion rates.(34) Moreover, its *d*-bands are furthest from the Fermi level (below −3 eV), allowing a wide energy range for the study of the chain's band structure (see Figure S1 in the Supporting Information (SI)).

In this work, we have overcome all the aforementioned obstacles and have generated an extended film of atomically precise zigzag chains on a vicinal Ag(111) surface, as evidenced by scanning tunneling microscopy (STM) and low energy electron diffraction (LEED). The electronic band structure of such films has been unraveled by means of ARPES and complemented by scanning tunneling spectroscopy (STS) measurements on Ag(111). In this way, we determine the experimental energy gap and visualize the spatial distribution of the frontier orbitals. Such wealth of experimental information has been clarified and expanded by a comprehensive set of density functional theory (DFT) calculations and electron plane wave expansion (EPWE) simulations.

**Results and Discussion**

We have produced a monolayer film of cross-conjugated zigzag chains from the surface polymerization of the 4,4″-dibromo-*meta* terphenyl (DMTP) molecular aromatic precursors via C–C coupling (see **Figure 1**a and **Methods** section). The template of choice is a vicinal Ag(111) crystal surface with linear, mono-atomic steps running parallel to the [11–2] direction(29) that corresponds to the so-called fully kinked (100% kinked) configuration of the step-edge. We used this particular substrate since it provides a higher flexibility to reconstruct and therefore accommodate the produced zigzag structures more efficiently (cf. **Methods** section and **Figure S1** in the SI). Indeed, we can already disclose that we achieved an excellent film featuring a high yield of well-ordered and aligned zigzag chains.

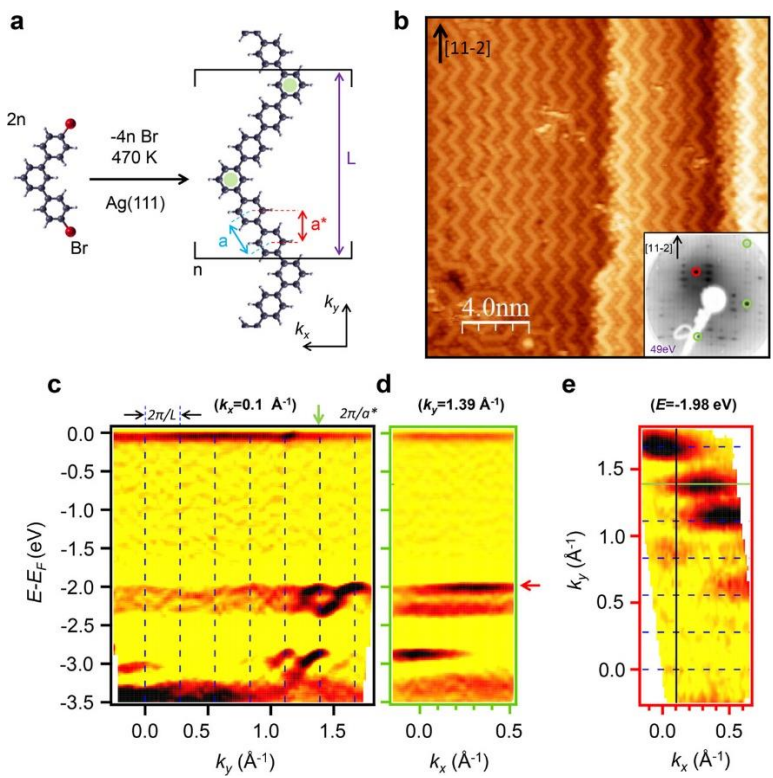

**Figure 1**. Structural arrangement and ARPES electronic band structure of the zigzag chain film grown on a vicinal Ag(111) surface. (a) Schematic representation of the DMTP precursor and the resulting zigzag covalent chain, showing its characteristic lengths: phenyl–phenyl distance (*a*) and its projection along the chain's average direction (*a**), and polymer superperiodicity (*L*). (b) High resolution STM image after chain synthesis on a vicinal plane ~3.6° off from the (111) crystal position. The zigzag chains are separated by Br atoms and preferentially follow the step parallel direction ([11–2]). (STM parameters: $V = -394$ mV, $I = 234$ pA, $T_{\text{sample}} = 100$ K). Inset shows the LEED pattern after chain formation that exhibits single-domain, well-aligned arrangement. The superstructure spots are in registry with the circled main spots (in red the (0,0) and in green the substrate's first order diffractions), implying commensurability with the terrace atoms (LEED parameters: $E_{\text{kin}} = 49$ eV, $T_{\text{sample}} = 300$ K). (c) ARPES experimental band structure of *meta*-junctioned cross-conjugated zigzag chains parallel to the average direction of chains and steps ($E$ vs $k_y$ with $k_x = 0.1$ Å$^{-1}$). (d) Experimental band structure perpendicular to the chain average axis ($E$ vs $k_x$, with $k_y = 1.39$ Å$^{-1}$ indicated by green arrow in (c)). (e) Isoenergetic cut ($k_x$ vs $k_y$) at the top of the valence molecular band ($E = -1.98$ eV, marked by red arrow in (d)). The second derivative of the intensity is shown in a linear color scale (highest being black). (ARPES parameters: $h\nu = 21.2$ eV, $T_{\text{sample}} = 150$ K).

The formed zigzag chains appear practically planar on the surface (Figure 1b) and are covalently bonded, displaying the characteristic phenyl–phenyl distance of $a \sim 4.3$ Å along the straight segments and a superperiodicity of $L \sim 2.24$ nm between equivalent elbows.(22,35) The unit cell of the chain features two straight subunits made up of two phenyl rings (in *para*-positions) linked to two edge rings acting as *meta*-junctions (Figure 1a). Note that in the STM image these chains are separated by spherical features attributed to Br atoms split off from the precursor molecules at the initial step of the on-surface reaction.(36−43) The LEED pattern reveals that the organic chains are aligned parallel to the steps and show long-range order as they are commensurate with the underlying substrate (Figures 1b and S2). Particularly, the main silver diffraction spots (red and green circles) are surrounded by a set of spots aligned along the average step direction yielding a (9, 5; 0, 4) superstructure.

Our STM and LEED structural results contain the required ingredients (atomic precision of the structure, defined alignment, long-range order and minimization of lateral interactions by Br adatom presence) to expect the existence of a defined and coherent electronic band structure from these chains. Figure 1c–e shows the second derivative (to enhance the details) of the ARPES spectral weight obtained from such a film saturating the surface (raw data is shown in Figure S3 in the SI). The resulting electronic structure in the direction parallel to the average step direction and the main axis of the zigzag chains ($E$ vs $k_y$ with $k_x = 0.1$ Å$^{-1}$) exhibits weakly dispersive bands between −1.8 and −3.5 eV, separated by an ∼0.6 eV gap (cf. Figure 1c). None of these ARPES features are observable on the pristine substrate (cf. Figure S1 in the SI). A closer inspection reveals that each one of them consists of a pair of antiphase oscillatory bands (Figures S3 and S4 in the SI). The spectral intensity peaks around $2\pi/a^*$, where $a^*$ represents the projected phenyl–phenyl distance along the average chain direction (Figure 1a), assuring its molecular origin.(44) The faint replicas with $2\pi/L$ periodicity (vertical dashed blue lines) stem from the zigzag chain superperiodicity $L$ (Figure 1a), in agreement with the STM data set.

The 1D nature of these zigzag chains is demonstrated by the lack of dispersion perpendicular to the average chain axis. Figure 1d shows a representative cut ($E$ vs $k_x$) across the center of the sixth Brillouin zone (green arrow at $k_y = 1.39$ Å$^{-1}$ in Figure 1c), where discrete flat bands can be observed. This confirms that they stem from different molecular orbitals of the zigzag polymer.(45) The nondispersive character at the top of the valence band (red arrow at −1.98 eV) can also be traced from the isoenergetic cut ($k_x$ vs $k_y$) shown in Figure 1e, where 1D polymer bands replicate at each Brillouin zone center, gaining intensity for the larger $k_y$ values. These photoemission intensity modulations have been simulated with the EPWE method,(46,47) which confirms that these features are neither affected by the templating Ag surface nor by the presence of Br atoms intercalated between the chains (see Methods section and Figure S4 in the SI). Indeed, we experimentally find that the presence of Br embedded in between the zigzag chains only causes a rigid shift of the molecular band structure by 200 ± 50 meV to higher energy, according to Figure S5 in the SI and in agreement with previous work.(24,33)

Our ARPES results suggest that the zigzag chains are largely decoupled from the metallic substrate since the observed molecular bands do not show signs of hybridization with the substrate in that energy window. Besides, the chains are semiconducting in nature with a band gap certainly larger than 2 eV, since no other bands closer to the Fermi energy are observed in the occupied region. The band structure strongly contrasts with that of the poly(*para*-phenylene) (called PPP hereafter) chains, which exhibits a single, highly dispersive molecular band across the entire Brillouin zone(23−25) (Figure S6 in the SI). Instead, it closely resembles the one predicted for poly(*meta*-phenylene) (called PMP hereafter) chains,(5) implying that the presence of *meta*-junctions strongly modifies the electronic structure of a polymeric chain(13) (cf. Figure S7 in the SI).

The weak interaction observed between the zigzag chain film and the substrate is a favorable playground for a systematic theoretical analysis. As a first approximation, we consider the polymers as free-standing and planar. On this basis, we use DFT calculations to corroborate the weakly dispersive band structure observed experimentally. The calculated electronic structure shown in Figure 2a exhibits convincing qualitative agreement with the experimental data. In particular, the dispersive character of the first four valence bands (VBs) of the zigzag chain (between −1 and −2.5 eV) is consistent with that in Figure 1c. The energy mismatch can be attributed to the absence of a substrate in the calculations, as well as to the well-known limitation of DFT to accurately predict HOMO–LUMO gaps. Note that the calculated bands span from the $\bar{\Gamma}$ point to the Brillouin zone boundary ($\pi/L$), which in the experiment appears 12 times replicated until $2\pi/a^*$. For comparison, the calculations are extended to straight PPP chains (Figure 2b) which strongly differ in the electronic structure by exhibiting a highly dispersive single VB in the same energy window. Moreover, the zigzag chain exhibits a greater band gap than its straight counterpart, confirming the enhanced semiconductive character of the former (see Figure S7 in the SI).

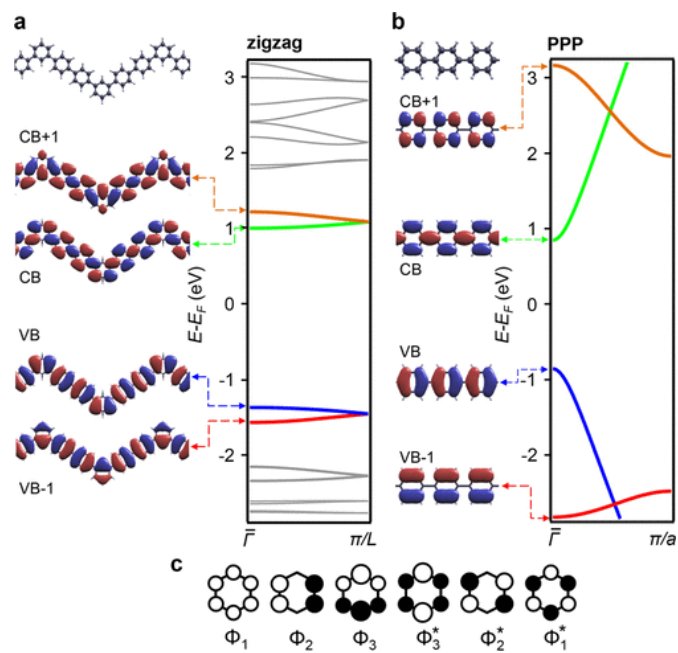

**Figure 2**. Comparison of molecular orbital shape and band structure between zigzag chains and straight PPP chains, as obtained from DFT calculations. The right plots in (a) and (b) show the calculated electronic band structure, where (a) corresponds to the zigzag polymer and (b) to PPP. The highly dispersive character of the PPP bands contrasts with the practically flat bands of the zigzag chains, accompanied by a notable difference in the frontier orbital band gap. Left panels in (a) and (b) show the spatially resolved molecular orbitals at $\bar{\Gamma}$ for each band. In a simplistic view, they are constructed by overlapping different benzene molecular orbitals, which are schematically shown in (c).

The experimental value of the frontier orbital band gap of the zigzag chains can be obtained by low-temperature (4 K) STS. For such measurements, we deposit a submonolayer coverage of DMTP molecules on Ag(111) so that small zigzag island patches are formed on the surface while still allowing access to the bare substrate for tip calibration and treatment (Figures 3 and S5). Figure 3a shows the d$I$/d$V$ spectra at the center (red) of a straight arm of a zigzag chain (see figure inset) and the Ag substrate (gray). The VB onset is detected close to −2.1 V (coinciding with the ARPES value in Figure 1d) while the conduction band (CB) edge is around 1.6 V resulting in an overall band gap of ∼3.7 V. Therefore, this value is larger than the 3.2 V reported for PPP chains grown on Au(111)[42] and confirms the enhanced semiconducting character of the zigzag chains.

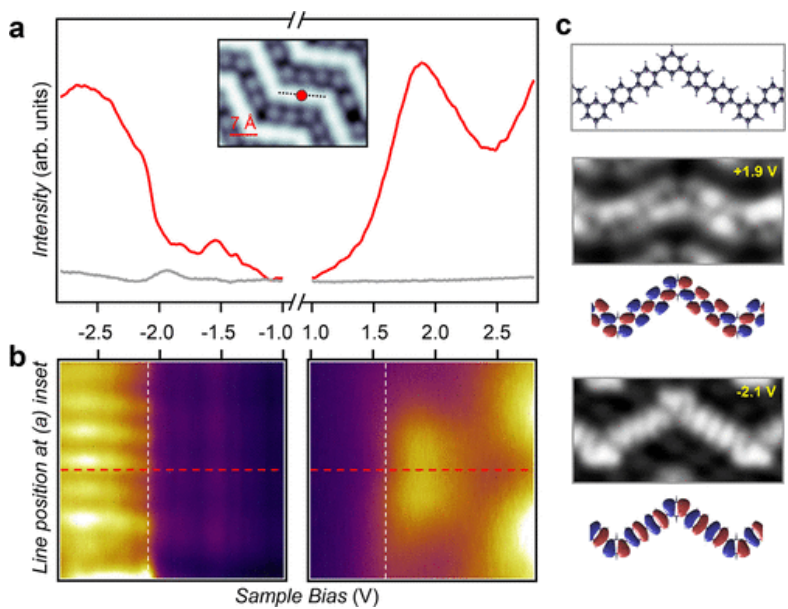

**Figure 3**. STS experimental determination of the zigzag chain's frontier orbitals on Ag(111). (a) Constant-height d$I$/d$V$ spectra acquired at the center of a zigzag straight arm (red point in STM image inset) and substrate (gray) (STM imaging parameters: 50 mV, 100 pA; frame: 3.6 × 2.5 nm$^2$). (b) Constant-height d$I$/d$V$ linescan spectra along the zigzag arm (indicated by a dotted line in the STM topography inset) for the same bias range of panel (a). The onsets of the VB and CB are clearly defined (vertical dashed white lines), yielding a band gap of ∼3.7 eV (STS bias voltage modulation for (a) and (b): 10 mV$_{rms}$ at 341 Hz. Close-feedback parameters: −350 mV, 150 pA and 1200 mV, 100 pA for the negative and positive resonances regions, respectively). (c) From top to bottom: ball and stick model of the zigzag chain. High-resolution d$I$/d$V$ maps acquired at constant-height with a CO functionalized STM tip at 1.9 and −2.1 V, *i.e.*, close to the CB and VB onsets (frames: 3.0 × 1.5 nm$^2$; bias voltage modulation 10 mV$_{rms}$ at 341 Hz). Underneath each map, the corresponding DFT gas phase molecular frontier orbitals are shown for comparison.

DFT calculations can also shed light onto the effect that the periodically spaced *meta*-junctions have on the overall electronic structure by comparing the spatially resolved molecular orbitals at the $\bar{\Gamma}$ point with the π molecular orbitals of benzene (Figure 2c). In the PPP case (Figure 2b), VB and CB are constructed by the overlap of $\Phi_3$ and $\Phi_3^*$ benzene molecular orbitals, respectively. These orbitals present a large electronic weight on the carbon atoms linking the phenyl rings (*para*-positions), giving rise to highly dispersive valence and conduction bands. Likewise, the less dispersive character of the VB-1 and CB+1 bands can be attributed to the orbital set that exhibits a nodal plane through the *para* carbon atoms ($\Phi_2$ and $\Phi_2^*$ orbitals). Contrarily, for the zigzag chains (Figure 2a) the VB and CB are a combination of two degenerate orbitals.(7) In particular, the VB is made up of $\Phi_3$ (straight sections) and $\Phi_2$ (elbows) orbitals, which is mirrored in the CB by $\Phi_3^*$ (straight sections) and $\Phi_2^*$ (elbows). This orbital mixing, along with the reduced orbital amplitude at the *meta*-positions and expected phase shifts induced by momentum steering at the elbows, results in a diminished orbital interaction (overlap) that leads to a severe weakening of the electron coupling between adjacent straight segments. Indeed, the flat band character is also exhibited by the VB-1 and CB+1, even though they mostly arise from a single type of benzene molecular orbital coupling ($\Phi_3$ and $\Phi_3^*$, respectively). This strong electronic effect governed by the *meta*-junction is generally referred to as cross-conjugation(5,6) or destructive quantum interference.(8−10,14)

The reduced electronic coupling between neighboring linear segments causes electron localization, an effect that can be adequately addressed with STS. Figure 3b presents a color plot representing stacked d$I$/d$V$ point spectra measured along a single straight segment (black dashed line in the inset of Figure 3a). Aside from clearly visualizing an overall band gap of 3.7 eV, we observe confinement in the CB within such segments where the spatial modulations in the local density of states (LDOS) are consistent with the first two stationary states of a particle in a box. In particular, their amplitudes die away at the edges of the linear segments (elbow positions of the zigzag chains) but the lower state at 1.9 V features an antinode at the segment's center (the peak in the red spectrum of Figure 3a), whereas the second state at 2.5 V oppositely exhibits a node at that position. Such electron confinement effects have also been observed in related structures, as in the case of finite size PPP chains featuring a single elbow (in *meta*-junction)(13) or in closed-cycle geometries of honeycombenes.(48) In essence, we can conclude that *meta*-junctions act as scattering barriers for the polymer electrons regardless of the overall geometry, *i.e.*, as closed structures(48) or as edged (nonlinear) chains.(13)

Once we have verified that each straight segment of our zigzag chains acts as a confining unit, reminiscent of a 1D array of weakly interacting Quantum Dots (QDs),(49) it should be possible to tune their electronic properties by modulating the straight segment's length. Varying the 1D QD length should affect the energy levels as well as the corresponding frontier orbital band gap. To do so, we coevaporated on Ag(111) linear precursors (DBTP molecules(24)) together with the previously used ones to generate the zigzag chains, as shown in Figure 1a. The Ullmann coupling reaction is likewise activated by postannealing to 470 K, resulting in linear segments of phenyl length configurations of $N = 4 + 3n$ (with $N$ being the total phenyl number and $n$ the amount of DBTP precursors embedded in the straight segment). Figure 4b shows color plots of the d$I$/d$V$ linescan for $N = 7$ (bottom) and $N = 10$ (top) QDs, which evidence the expected squared wave function intensity variations in the CBs for the same energy range as Figure 3b. Most importantly, by comparison to the dashed white lines corresponding to the $N = 4$ segment, we observe that the band gap shrinks as the size of the segment increases (dashed blue lines). A quantitative analysis of the experimentally determined band gap is shown in Figure 4c, revealing a 1/$N$ behavior in agreement with previous work for similar chains.(13) Such behavior matches our DFT calculations for planar, freestanding, periodic zigzag chains of different straight segment length, which confirm not only that the band gap of the cross-conjugated zigzag chains is larger than the one of its linear PPP counterpart, but it is also tunable as 1/$N$ (cf. Figures S7 and S8 in the SI).

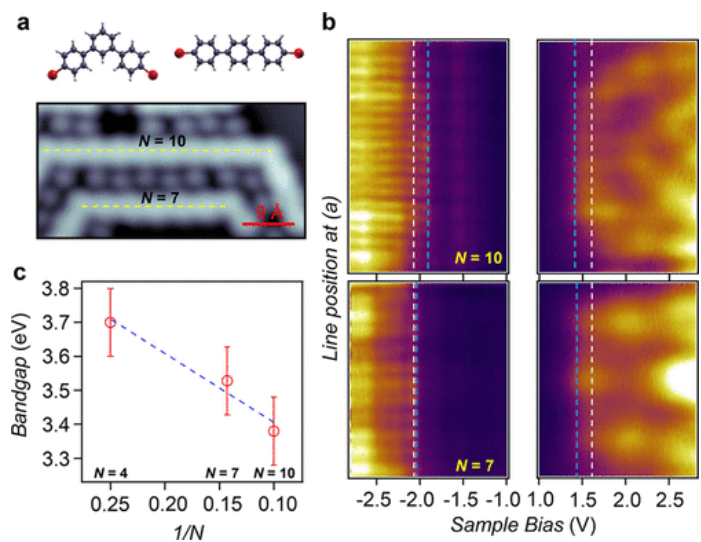

**Figure 4**. Tuning the electronic confined states through the linear segment's length (QD size). (a) Schematic representation of the coevaporated molecular precursors (DMTP and DBTP) that generate longer straight segments between the *meta*-coordinating phenyls. The STM image shows two longer linear segments of 7 and 10 phenyl rings between elbows (Imaging parameters: 50 mV, 100 pA; frame: 4.5 × 2.3 nm$^2$). (b) Constant-height d$I$/d$V$ line profiles close to the VB and CB onsets along the two segments following the dashed lines in (a). The intensity modulation of the CB confirms the confinement nature of the *meta*-junction termination. Furthermore, the frontier orbital band gap (vertical dashed blue lines) is reduced compared to the $N = 4$ case, which is indicated by the two vertical dashed white lines. (STS parameters: bias voltage modulation of 10 mV$_{rms}$ at 341 Hz; close-feedback at −350 mV, 150 pA and 1200 mV, 120 pA for the negative and positive resonances regions, respectively). (c) Experimental band gap extracted from the STS line profiles revealing a 1/$N$ behavior (compare to Figure S8 in the SI).

Finally, we should discuss an additional property of this system that may affect the frontier orbital band gap: the relative twisting of the phenyl rings (nonplanar chain morphology). Figure 3c shows high-resolution constant height d$I$/d$V$ maps close to the valence and conduction band onsets, namely, at −2.1 and 1.9 eV, respectively. Both maps replicate well the molecular orbital simulations of Figure 2a, which are calculated for planar structures and are depicted below for direct comparison. While the slight discrepancies in nodal positions are attributed to the CO probe functionalization,(50) the intensity variations are in turn ascribed to the twisting of the phenyl rings.(42) The phenyl twisting is confirmed from constant height bond resolution imaging with a CO functionalized tip,(51) on which we find subtle intensity modulations on the external parts of the phenyl rings (see Figure S9 in the SI) indicative of such phenyl twisting. This is in agreement with previous work that found ∼20° phenyl twisting on PPP chains with respect to the surface plane.(24) To determine the effect of the twisting on the frontier orbital band gap, we have performed DFT calculations for free-standing zigzag chains and obtained an energy minimum close to 30° phenyl twisting of the zigzag chain. However, the calculations show that the electronic band gap is scarcely increased compared to the planar configuration. Given that the underlying substrate is known to decrease the phenyl twisting by flattening the chains, we expect this effect to be smaller than ∼200 meV (cf. Figure S10 in the SI), validating the use of free-standing, planar configuration geometries in the DFT calculations that support our findings.

**Conclusions**

We have been able to synthesize and macroscopically align a saturated film of cross-conjugated oligophenylene zigzag chains on a vicinal Ag(111) surface. We find that these atomically precise chains remain sufficiently decoupled from each other and from the substrate to probe their elusive band structure with ARPES, revealing weakly dispersing one-dimensional electronic bands along the chain direction. DFT band structure calculations and EPWE photoemission intensity simulations satisfactorily reproduce our experimental findings. By means of STS, we find that the zigzag chain has a larger frontier orbital band gap than its straight counterpart (PPP) and observe electronic confinement in each straight segment of the zigzag chains, reminiscent of 1D arrays of weakly interacting QDs. Such states can be tuned by changing the length of the straight segments, affecting the frontier orbital band gap, which follows a 1/$N$ dependency. Indeed, our molecular orbital simulations confirm that the periodically spaced *meta*-junctions at the elbows of the zigzag chain are the main structural feature responsible for the reduction of electronic coupling between adjacent linear segments. These findings corroborate the important effects that the conductive path topology of a molecular wire has on its frontier orbitals, which are responsible for defining its chemical, optical and electronic properties. Recent advances in transfer techniques ensure that on-surface synthesized and well-aligned organic nanostructures can be collectively transferred onto insulating substrates maintaining their relative arrangement,(52−54) which opens the path to further study the transport and optical properties of these cross-conjugated oligophenylene zigzag chains.

**Methods**

A silver crystal surface curved around the (645) direction was used as tunable vicinal substrate for chain formation and alignment.(29) This curved sample exhibits (111) terraces of variable size (position dependent on its curvature) separated by monoatomic steps oriented along the [11–2] direction. The steps are of fully kinked type, in which out-protruding atoms have no side neighbors (Figure S1 in the SI). Notably, periodic roughening of such step-edges bears negligible energy cost, and hence can readily accommodate to the zigzag structure of the chains.(29) The saturated zigzag film that presents best order was observed at the vicinal plane ∼ 3.6° from the (111) region, corresponding to an average terrace size of 3.8 nm. The ARPES data shown in Figures 1, S1, S3, S4, and S6 corresponds to this position of the substrate. The sample was cleaned by repeated cycles of Ar$^+$ sputtering at energies of 1.0 keV, followed by annealing at 700 K. This produced clean and well-ordered surface step arrays as verified by the splitting of the LEED spots along the surface curvature.

The *meta*-junctioned haloaromatic compound, DMTP, was sublimated from a Knudsen cell at 360 K at a low flux (1 ML/hour) while the sample was held at 470 K.(22,48,55) Covalently bonded zigzag chains appeared separated by Br adatoms, suggesting some influence from the latter in steering chain growth and alignment. The step flexibility promoted the chain formation while keeping other irregular structures or hyperbenzene macrocycles to a minimum.(41) The structures remained densely packed up to high temperatures (~600 K), beyond which halogen desorption starts, accompanied by chain misalignment.(18,56)

ARPES measurements were performed with a lab-based experimental setup using a display-type hemispherical electron analyzer (SPECS Phoibos 150, energy/angle resolution of 40 meV/0.1°) combined with a monochromatized Helium I ($h\nu$ = 21.2 eV) source. Measurements were acquired with the sample at 150 K by moving the polar angle, which is set to be parallel to the average step direction.

The STM measurements on the curved Ag(111) crystal were carried out at 100 K using a variable temperature Omicron STM instrument with a Nanonis SPM control system. The bias voltages given in the paper refer to a grounded tip. STM data were acquired in constant current mode and were processed with the WSxM software.(57)

LT-STM/STS measurements were performed with a commercial Scienta-Omicron low temperature system, operating in ultrahigh vacuum (UHV) at 4.3 K. For the measurement, the bias voltage was applied to the tip while the sample was electronically grounded. The STM tip was prepared *ex situ* by clipping a Pt/Ir wire (0.25 mm) and sharpened *in situ* by repeatedly indenting the tip a few nanometers (1–4 nm) into the Ag surface while applying bias voltages from 2 to 4 V between tip and sample. In order to perform bond-resolved STM imaging, the tip apex was terminated with a CO molecule, directly picked up from the surface, by positioning the sharp metal tip on top of it and applying a 500 ms bias pulse at −2 V. The imaging was performed by measuring at constant height while applying a bias voltage to the tip within the range of 2.0 and 3.5 mV. For spectroscopic point spectra and conductance maps, the d$I$/d$V$ signals were measured by a digital lock-in amplifier (Nanonis). STM images were analyzed by using the WSxM software.(57)

*Ab initio* calculations were carried out using density functional theory, as implemented in the SIESTA code.(58) The optB88-vdW functional,(59) which accounts for nonlocal corrections, was adopted for the exchange and correlation potential. For each organic nanostructure, we considered a supercell consisting of a chain infinite along the $x$ axis, with vacuum gaps of 15 Å in $y$ and $z$ directions in order to avoid interactions between chains in adjacent cells. A Monkhorst–Pack k-point grid with 20 × 1 × 1 k-points was used for the Brillouin zone sampling and the mesh cutoff for real space integrations was set to 300 Ry. We employed a double-ξ plus polarization (DZP) basis set, and a mesh-cutoff of 300 Ry for the real-space integrations. All structures were fully relaxed until residual forces were less than 0.01 eV/Å.

The electron plane wave expansion method, which was recently used to describe the electronic properties of graphene nanostructures,(47) is employed to simulate ARPES data. Within the EPWE approach, the photoemission intensity for a given binding energy and photoelectron wave vector is obtained from Fermi's golden rule applied to the in-plane wave function (an initial state) and a normalized plane wave (a final state) for the parallel component of the photoelectron wave function, as detailed in ref. (47). In this semiempirical method, zigzag chains are considered free-standing and planar, which implies that the simulated bands are substrate independent and free of Br interactions.

### Acknowledgements

I.P.-Z. thanks Dr. Jens Brede for fruitful discussion. Z.M.A. thanks Prof. F. J. García de Abajo for providing the EPWE code. We acknowledge the financial support from the Spanish Ministry of Economy, Industry and Competitiveness (MINECO, Grant No. MAT2016-78293-C6), from the Basque Government (Grant No. IT-621-13), from the regional Government of Aragon (RASMIA project), from the European Regional Development Fund (ERDF) under the program Interreg V-A España-Francia-Andorra (Contract No. EFA 194/16 TNSI), from the German Science Foundation (DFG) through the CRC 1083 and Grant

No. GO 1812/2 and funding from the European Research Council (ERC) under the European Union's Horizon 2020 research and innovation programme (grant agreement no. 635919).

**References**

(1) Günes, S.; Neugebauer, H.; Sariciftci, N. S. Conjugated Polymer-Based Organic Solar Cells. Chem. Rev. 2007, 107, 1324−1338.

(2) Li, G.; Chang, W.-H.; Yang, Y. Low-Bandgap Conjugated Polymers Enabling Solution Processable Tandem Solar Cells. Nat. Rev. Mater. 2017, 2, 17043.

(3) Masai, H.; Terao, J. Stimuli-Responsive Functionalized Insulated Conjugated Polymers. Polym. J. 2017, 49, 805−814.

(4) Guiglion, P.; Zwijnenburg, M. A. Contrasting the Optical Properties of the Different Isomers of Oligophenylene. Phys. Chem. Chem. Phys. 2015, 17, 17854−17863.

(5) Hong, S. Y.; Kim, D. Y.; Kim, C. Y.; Hoffmann, R. Origin of the Broken Conjugation in m-Phenylene Linked Conjugated Polymers. Macromolecules 2001, 34, 6474−6481.

(6) van der Veen, M.; Rispens, M.; Jonkman, H.; Hummelen, J. Molecules with Linear π- Conjugated Pathways between All Substituents: Omniconjugation. Adv. Funct. Mater. 2004, 14, 215−223.

(7) Kocherzhenko, A. A.; Grozema, F. C.; Siebbeles, L. D. A. Single Molecule Charge Transport: From a Quantum Mechanical to a Classical Description. Phys. Chem. Chem. Phys. 2011, 13, 2096−2110.

(8) Manrique, D. Z.; Huang, C.; Baghernejad, M.; Zhao, X.; Al-Owaedi, O. A.; Sadeghi, H.; Kaliginedi, V.; Hong, W.; Gulcur, M.; Wandlowski, T.; Bryce, M. R.; Lambert, C. J. A Quantum Circuit Rule for Interference Effects in Single-Molecule Electrical Junctions. Nat. Commun. 2015, 6, 6389.

(9) Guédon, C. M.; Valkenier, H.; Markussen, T.; Thygesen, K. S.; Hummelen, J. C.; van der Molen, S. J. Observation of Quantum Interference in Molecular Charge Transport. Nat. Nanotechnol. 2012, 7, 305−309.

(10) Markussen, T.; Stadler, R.; Thygesen, K. S. The Relation between Structure and Quantum Interference in Single Molecule Junctions. Nano Lett. 2010, 10, 4260−4265.

(11) Limacher, P. A.; Lüthi, H. P. Cross-Conjugation. Wiley Interdiscip. Rev. Comput. Mol. Sci. 2011, 1, 477−486.

(12) Thompson, A. L.; Ahn, T.-S.; Thomas, K. R. J.; Thayumanavan, S.; Martínez, T. J.; Bardeen, C. J. Using Meta Conjugation To Enhance Charge Separation versus Charge Recombination in Phenylacetylene Donor-Bridge-Acceptor Complexes. J. Am. Chem. Soc. 2005, 127, 16348−16349.

(13) Wang, S.; Wang, W.; Lin, N. Resolving Band-Structure Evolution and Defect-Induced States of Single Conjugated Oligomers by Scanning Tunneling Microscopy and Tight-Binding Calculations. Phys. Rev. Lett. 2011, 106, 206803.

(14) Tada, T.; Yoshizawa, K. Molecular Design of Electron Transport with Orbital Rule: Toward Conductance-Decay Free Molecular Junctions. Phys. Chem. Chem. Phys. 2015, 17, 32099−32110.

(15) Terao, J.; Wadahama, A.; Matono, A.; Tada, T.; Watanabe, S.; Seki, S.; Fujihara, T.; Tsuji, Y. Design Principle for Increasing Charge Mobility of π-Conjugated Polymers Using Regularly Localized Molecular Orbitals. Nat. Commun. 2013, 4, 1691.

(16) Cai, J.; Ruffieux, P.; Jaafar, R.; Bieri, M.; Braun, T.; Blankenburg, S.; Muoth, M.; Seitsonen, A. P.; Saleh, M.; Feng, X.; Müllen, K.; Fasel, R. Atomically Precise Bottom-Up Fabrication of Graphene Nanoribbons. Nature 2010, 466, 470−473.

(17) Talirz, L.; Ruffieux, P.; Fasel, R. On-Surface Synthesis of Atomically Precise Graphene Nanoribbons. Adv. Mater. 2016, 28, 6222−6231.

(18) de Oteyza, D. G.; Garcia-Lekue, A.; Vilas-Varela, M.; Merino-Díez, N.; Carbonell-Sanromà, E.; Corso, M.; Vasseur, G.; Rogero, C.; Guitián, E.; Pascual, J. I.; Ortega, J. E.; Wakayama, Y.; Peña, D. Substrate-Independent Growth of Atomically Precise Chiral Graphene Nanoribbons. ACS Nano 2016, 10, 9000−9008.

(19) Corso, M.; Carbonell-Sanromà, E.; de Oteyza, D. G. In On-Surface Synthesis II; de Oteyza, D. G., Rogero, C., Eds.; Springer International Publishing: Cham, 2018; pp 113−152.

(20) Gröning, O.; Wang, S.; Yao, X.; Pignedoli, C. A.; Borin Barin, G.; Daniels, C.; Cupo, A.; Meunier, V.; Feng, X.; Narita, A.; Müllen, K.; Ruffieux, P.; Fasel, R. Engineering of Robust Topological Quantum Phases in Graphene Nanoribbons. Nature 2018, 560, 209−213.

(21) Rizzo, D. J.; Veber, G.; Cao, T.; Bronner, C.; Chen, T.; Zhao, F.; Rodriguez, H.; Louie, S. G.; Crommie, M. F.; Fischer, F. R. Topological Band Engineering of Graphene Nanoribbons. Nature 2018, 560, 204−208.

(22) Fan, Q.; Wang, C.; Han, Y.; Zhu, J.; Hieringer, W.; Kuttner, J.; Hilt, G.; Gottfried, J. M. Surface-Assisted Organic Synthesis of Hyperbenzene Nanotroughs. Angew. Chem., Int. Ed. 2013, 52, 4668−4672.

(23) Vasseur, G.; Fagot-Revurat, Y.; Sicot, M.; Kierren, B.; Moreau, L.; Malterre, D.; Cardenas, L.; Galeotti, G.; Lipton-Duffin, J.; Rosei, F.; Di Giovannantonio, M.; Contini, G.; Le Fèvre, P.; Bertran, F.; Liang, L.; Meunier, V.; Perepichka, D. F. Quasi One-Dimensional Band Dispersion and Surface Metallization in Long-Range Ordered Polymeric Wires. Nat. Commun. 2016, 7, 10235.

(24) Basagni, A.; Vasseur, G.; Pignedoli, C. A.; Vilas-Varela, M.; Peña, D.; Nicolas, L.; Vitali, L.; Lobo-Checa, J.; de Oteyza, D. G.; Sedona, F.; Casarin, M.; Ortega, J. E.; Sambi, M. Tunable Band Alignment with Unperturbed Carrier Mobility of On-Surface Synthesized Organic Semiconducting Wires. ACS Nano 2016, 10, 2644−2651.

(25) Abadía, M.; Ilyn, M.; Piquero-Zulaica, I.; Gargiani, P.; Rogero, C.; Ortega, J. E.; Brede, J. Polymerization of Well-Aligned Organic Nanowires on a Ferromagnetic Rare-Earth Surface Alloy. ACS Nano 2017, 11, 12392−12401.

(26) Ortega, J. E.; Mugarza, A.; Repain, V.; Rousset, S.; Pérez-Dieste, V.; Mascaraque, A. One-Dimensional versus Two-Dimensional Surface States on Stepped Au(111). Phys. Rev. B: Condens. Matter Mater. Phys. 2002, 65, 165413.

(27) Mugarza, A.; Schiller, F.; Kuntze, J.; Cordón, J.; Ruiz-Osés, M.; Ortega, J. E. Modelling Nanostructures with Vicinal Surfaces. J. Phys.: Condens. Matter 2006, 18, S27.

(28) Corso, M.; Schiller, F.; Fernández, L.; Cordón, J.; Ortega, J. E. Electronic States in Faceted Au(111) Studied with Curved Crystal Surfaces. J. Phys.: Condens. Matter 2009, 21, 353001.

(29) Ortega, J. E.; Vasseur, G.; Piquero-Zulaica, I.; Matencio, S.; Valbuena, M. A.; Rault, J. E.; Schiller, F.; Corso, M.; Mugarza, A.; Lobo-Checa, J. Structure and Electronic States of Vicinal Ag(111) Surfaces with Densely Kinked Steps. New J. Phys. 2018, 20, 073010.

(30) Ruffieux, P.; Cai, J.; Plumb, N. C.; Patthey, L.; Prezzi, D.; Ferretti, A.; Molinari, E.; Feng, X.; Müllen, K.; Pignedoli, C. A.; Fasel, R. Electronic Structure of Atomically Precise Graphene Nanoribbons. ACS Nano 2012, 6, 6930−6935.

(31) Senkovskiy, B. V.; Usachov, D. Y.; Fedorov, A. V.; Haberer, D.; Ehlen, N.; Fischer, F. R.; Grüneis, A. Finding the Hidden Valence Band of N = 7 Armchair Graphene Nanoribbons with Angle-Resolved Photoemission Spectroscopy. 2D Mater. 2018, 5, 035007.

(32) Cai, L.; Yu, X.; Liu, M.; Sun, Q.; Bao, M.; Zha, Z.; Pan, J.; Ma, H.; Ju, H.; Hu, S.; Xu, L.; Zou, J.; Yuan, C.; Jacob, T.; Björk, J.; Zhu, J.; Qiu, X.; Xu, W. Direct Formation of C−C Double-Bonded Structural Motifs by On-Surface Dehalogenative Homocoupling of gem-Dibromomethyl Molecules. ACS Nano 2018, 12, 7959−7966.

(33) Merino-Díez, N.; Lobo-Checa, J.; Nita, P.; Garcia-Lekue, A.; Basagni, A.; Vasseur, G.; Tiso, F.; Sedona, F.; Das, P. K.; Fujii, J.; Vobornik, I.; Sambi, M.; Pascual, J. I.; Ortega, J. E.; de Oteyza, D. G. Switching from Reactant to Substrate Engineering in the Selective Synthesis of Graphene Nanoribbons. J. Phys. Chem. Lett. 2018, 9, 2510−2517.

(34) Bieri, M.; Nguyen, M.-T.; Gröning, O.; Cai, J.; Treier, M.; Aït-Mansour, K.; Ruffieux, P.; Pignedoli, C. A.; Passerone, D.; Kastler, M.; Müllen, K.; Fasel, R. Two-Dimensional Polymer Formation on Surfaces: Insight into the Roles of Precursor Mobility and Reactivity. J. Am. Chem. Soc. 2010, 132, 16669−16676.

(35) Judd, C. J.; Haddow, S. L.; Champness, N. R.; Saywell, A. Ullmann Coupling Reactions on Ag(111) and Ag(110); Substrate Inuence on the Formation of Covalently Coupled Products and Intermediate Metal-Organic Structures. Sci. Rep. 2017, 7, 14541.

(36) Wang, W.; Shi, X.; Wang, S.; Van Hove, M. A.; Lin, N. Single-Molecule Resolution of an Organometallic Intermediate in a Surface-Supported Ullmann Coupling Reaction. J. Am. Chem. Soc. 2011, 133, 13264−13267.

(37) Di Giovannantonio, M.; El Garah, M.; Lipton-Duffin, J.; Meunier, V.; Cardenas, L.; Fagot Revurat, Y.; Cossaro, A.; Verdini, A.; Perepichka, D. F.; Rosei, F.; Contini, G. Insight into Organometallic Intermediate and Its Evolution to Covalent Bonding in Surface-Confined Ullmann Polymerization. ACS Nano 2013, 7, 8190−8198.

(38) Fan, Q.; Wang, C.; Han, Y.; Zhu, J.; Kuttner, J.; Hilt, G.; Gottfried, J. M. Surface Assisted Formation, Assembly, and Dynamics of Planar Organometallic Macrocycles and Zigzag Shaped Polymer Chains with C-Cu-C Bonds. ACS Nano 2014, 8, 709−718.

(39) Koch, M.; Gille, M.; Viertel, A.; Hecht, S.; Grill, L. Substrate-Controlled Linking of Molecular Building Blocks: Au(111) vs. Cu(111). Surf. Sci. 2014, 627, 70−74.

(40) Chen, M.; Shang, J.; Wang, Y.; Wu, K.; Kuttner, J.; Hilt, G.; Hieringer, W.; Gottfried, J. M. On-Surface Synthesis and Character­ization of Honeycombene Oligophenylene Macrocycles. ACS Nano 2017, 11, 134−143.

(41) Fan, Q.; Wang, T.; Dai, J.; Kuttner, J.; Hilt, G.; Gottfried, J. M.; Zhu, J. On-Surface Pseudo-High-Dilution Synthesis of Macrocycles: Principle and Mechanism. ACS Nano 2017, 11, 5070−5079.

(42) Merino-Díez, N.; Garcia-Lekue, A.; Carbonell-Sanromà, E.; Li, J.; Corso, M.; Colazzo, L.; Sedona, F.; Sánchez-Portal, D.; Pascual, J. I.; de Oteyza, D. G. Width Dependent Band Gap in Armchair Graphene Nanoribbons Reveals Fermi Level Pinning on Au(111). ACS Nano 2017, 11, 11661−11668.

(43) Fan, Q.; Liu, L.; Dai, J.; Wang, T.; Ju, H.; Zhao, J.; Kuttner, J.; Hilt, G.; Gottfried, J. M.; Zhu, J. Surface Adatom Mediated Structural Transformation in Bromoarene Monolayers: Precursor Phases in Surface Ullmann Reaction. ACS Nano 2018, 12, 2267−2274.

(44) Offenbacher, H.; Lüftner, D.; Ules, T.; Reinisch, E. M.; Koller, G.; Puschnig, P.; Ramsey, M. G. Orbital Tomography: Molecular Band Maps, Momentum Maps and the Imaging of Real Space Orbitals of Adsorbed Molecules. J. Electron Spectrosc. Relat. Phenom. 2015, 204, 92−101.

(45) Koller, G.; Berkebile, S.; Oehzelt, M.; Puschnig, P.; Ambrosch-Draxl, C.; Netzer, F. P.; Ramsey, M. G. Intra- and Intermolecular Band Dispersion in an Organic Crystal. Science 2007, 317, 351−355.

(46) Mugarza, A.; Mascaraque, A.; Pérez-Dieste, V.; Repain, V.; Rousset, S.; García de Abajo, F. J.; Ortega, J. E. Electron Connement in Surface States on a Stepped Gold Surface Revealed by Angle-Resolved Photoemission. Phys. Rev. Lett. 2001, 87, 107601.

(47) Abd El-Fattah, Z. M.; Kher-Elden, M. A.; Piquero-Zulaica, I.; García de Abajo, F. J.; Ortega, J. E. Graphene: Free Electron Scattering Within an Inverted Honeycomb Lattice. arXiv:1808.06034.

(48) Fan, Q.; Dai, J.; Wang, T.; Kuttner, J.; Hilt, G.; Gottfried, J. M.; Zhu, J. Confined Synthesis of Organometallic Chains and Macro-cycles by Cu-O Surface Templating. ACS Nano 2016, 10, 3747−3754.

(49) Piquero-Zulaica, I.; Lobo-Checa, J.; Sadeghi, A.; Abd El-Fattah, Z. M.; Mitsui, C.; Okamoto, T.; Pawlak, R.; Meier, T.; Arnau, A.; Ortega, J. E.; Takeya, J.; Goedecker, S.; Meyer, E.; Kawai, S. Precise Engineering of Quantum Dot Array Coupling Through Their Barrier Widths. Nat. Commun. 2017, 8, 787.

(50) Gross, L.; Moll, N.; Mohn, F.; Curioni, A.; Meyer, G.; Hanke, F.; Persson, M. High-Resolution Molecular Orbital Imaging Using a p-Wave STM Tip. Phys. Rev. Lett. 2011, 107, No. 086101, DOI: 10.1103/PhysRevLett.107.086101.

(51) Hieulle, J.; Carbonell-Sanromà, E.; Vilas-Varela, M.; Garcia-Lekue, A.; Guitián, E.; Peña, D.; Pascual, J. I. On-Surface Route for Producing Planar Nanographenes with Azulene Moieties. Nano Lett. 2018, 18, 418−423.

(52) Ohtomo, M.; Sekine, Y.; Hibino, H.; Yamamoto, H. Graphene Nanoribbon Field-Effect Transistors Fabricated by Etchant-Free Transfer from Au(788). Appl. Phys. Lett. 2018, 112, 021602.

(53) Llinas, J. P.; et al. Short-Channel Field-Effect Transistors with 9-Atom and 13-Atom Wide Graphene Nanoribbons. Nat. Commun. 2017, 8, 633.

(54) Moreno, C.; Vilas-Varela, M.; Kretz, B.; Garcia-Lekue, A.; Costache, M. V.; Paradinas, M.; Panighel, M.; Ceballos, G.; Valenzuela, S. O.; Peña, D.; Mugarza, A. Bottom Up Synthesis of Multifunctional Nanoporous Graphene. Science 2018, 360, 199−203.

(55) Fan, Q.; Gottfried, J. M.; Zhu, J. Surface-Catalyzed C-C Covalent Coupling Strategies toward the Synthesis of Low-Dimensional Carbon-Based Nanostructures. Acc. Chem. Res. 2015, 48, 2484−2494.

(56) Goddard, P.; Schwaha, K.; Lambert, R. Adsorption-Desorption Properties and Surface Structural Chemistry of Bromine on Clean and Sodium-Dosed Ag(111). Surf. Sci. 1978, 71, 351−363.

(57) Horcas, I.; Fernández, R.; Gómez-Rodríguez, J. M.; Colchero, J.; Gómez-Herrero, J.; Baro, A. M. WSXM: A Software for Scanning Probe Microscopy and a Tool for Nanotechnology. Rev. Sci. Instrum. 2007, 78, 013705.

(58) Soler, J. M.; Artacho, E.; Gale, J. D.; García, A.; Junquera, J.; Ordejón, P.; Sánchez Portal, D. The SIESTA Method for Ab Initio Order-N Materials Simulation. J. Phys.: Condens. Matter 2002, 14, 2745−2779.

(59) Klimes, J.; Bowler, D. R.; Michaelides, A. Chemical Accuracy for the Van Der Waals Density Functional. J. Phys.: Condens. Matter 2010, 22, 022201.

# Supplementary information of:

Electronic structure tunability by periodic *meta*-ligand spacing in one-dimensional organic semiconductors

Ignacio Piquero-Zulaica[*,1], Aran Garcia-Lekue[2,3], Luciano Colazzo[2], Claudio K. Krug[4], Mohammed Sabri[2], Zakaria M. Abd El-Fattah[5,6], J. Michael Gottfried[4], Dimas G. de Oteyza[1,2,3], J. Enrique Ortega[1,2,7], Jorge Lobo-Checa[*,8,9]

[1]Centro de Física de Materiales CSIC/UPV-EHU-Materials Physics Center, Paseo Manuel Lardizabal 5, E-20018 San Sebastián, Spain
[2]Donostia International Physics Center (DIPC), Paseo Manuel Lardizabal 4, E-20018 Donostia-San Sebastián, Spain
[3]Ikerbasque, Basque Foundation for Science, 48011 Bilbao, Spain
[4]Fachbereich Chemie, Philipps-Universität Marburg, Hans-Meerwein-Str. 4, 35032 Marburg, Germany
[5] ICFO-Institut de Ciencies Fotoniques, The Barcelona Institute of Science and Technology, 08860 Castelldefels, Barcelona, Spain
[6] Physics Department, Faculty of Science, Al-Azhar University, Nasr City, E-11884 Cairo, Egypt
[7]Dpto. Física Aplicada I, Universidad del País Vasco, E-20018 San Sebastián, Spain
[8]Instituto de Ciencia de Materiales de Aragón (ICMA), CSIC-Universidad de Zaragoza, E-50009 Zaragoza, Spain
[9]Departamento de Física de la Materia Condensada, Universidad de Zaragoza, E-50009 Zaragoza, Spain

This file discussess the following topics related to the main manuscript:

1. Growth details and zigzag chain alignment on a Ag(111) curved vicinal crystal.
2. ARPES band structure and photoemission intensity simulations using the Plane Wave Expansion (PWE) method.
3. Br effect on the electronic structure of zigzag chains studied by STM/STS.
4. ARPES band structure comparison of zigzag chains vs poly-(para-phenylene) (PPP) chains.
5. Band structure variations with straight segment's length of the zigzag chains from DFT calculations.
6. Phenyl twisting considerations of zigzag chains by STM/STS and DFT.

It contains 10 supporting figures and its corresponding references.

## I. Growth details and zigzag chain alignment on a Ag(111) curved vicinal crystal

We have used a curved vicinal Ag(111) crystal[1] featuring 100% kinked steps running along the [11-2] direction (Fig. S1a,b). These steps promote the long-range zigzag chain ordering, an essential requirement to obtain properly defined band structures. The substrate's band structure has the *d*-bands' onset below -3 eV, which provides a large energy window for the zigzag chain bands undisrupted observation (Fig. S1c,d).

Zigzag chains grow into a long-range ordered polymer film parallel to the steps at the sample position of ~3.6⁰ off from the (111) region, yielding a (9, 5; 0, 4) matrix superstructure in LEED (Figures S2 (a-c)). However, in the (111) region, the chains condense into multidomains, as extracted from STM topography images and macroscopically by LEED (Figures S2 (d-f)).

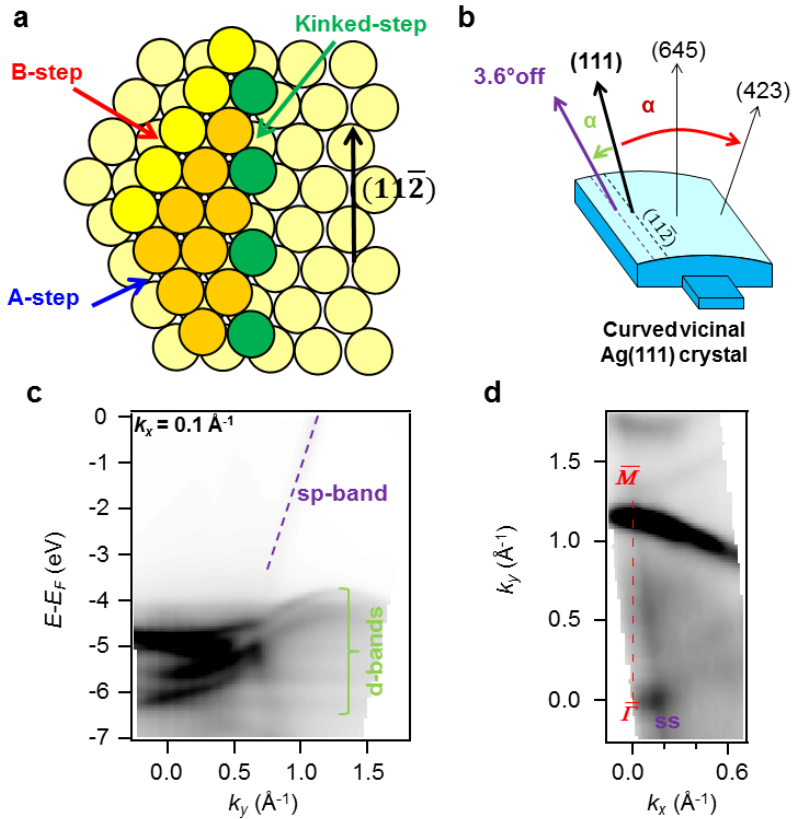

***Figure S1│ Atomic and macroscopic geometry of the 100% kinked curved Ag(111) crystal and its electronic structure ~3.6° off the (111) plane. (a)** Schematic drawing highlighting the three step types present on close-packed vicinal surfaces. In our case, the step termination corresponds to the one ending in the green spheres (100% kinked) that runs along the [11-2] direction. **(b)** Schematic drawing of the curved Ag(111) substrate. The crystal was polished so that the (111) region is away from the center of the crystal by 9.27°. Zigzag chain films showed best alignment at the position with a local vicinal plane of ~3.6° off from the (111) region, where kinked steps are 3.8 nm separated in average from each other. **(c)** The ARPES band structure (E vs $k_y$) at $k_x = 0.1$ Å$^{-1}$ corresponds to the direction parallel to the [11-2]. The characteristic silver d-bands exist below -3 eV and the sp-band shows a large dispersive character as it raises and crosses the Fermi level (highlighted by a side dashed purple line). **(d)** Fermi Surface Map ($k_x$ vs $k_y$ at E = 0), exhibiting the characteristic Shockley state (surface state) close to the $\bar{\Gamma}$ point and the dispersive dominant sp-bulk bands closer to the $\bar{M}$ point.*

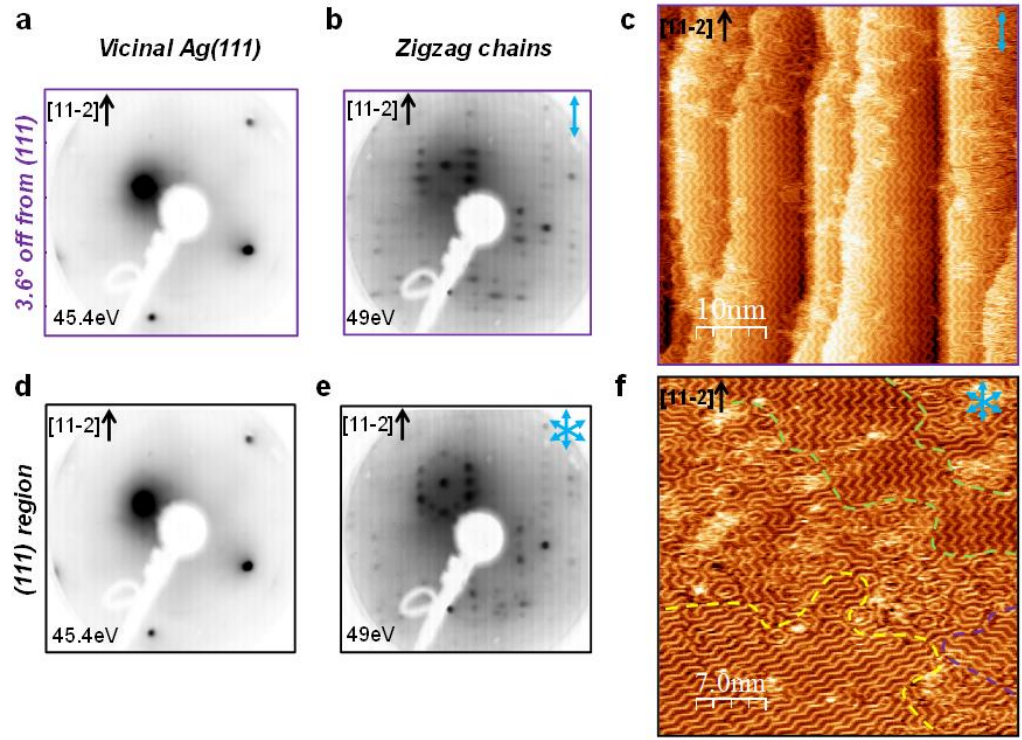

***Figure S2 | Uniaxial zigzag chain ordering vs multi-domain structures determined by LEED and STM.*** Experimental LEED patterns of the silver substrate before (**a**) and after (**b**) the formation of the zigzag chain film at the ~3.6° position off from the (111). This pattern agrees with a (9, 5; 0, 4) matrix, which corresponds to an elbow-elbow distance of 9 atomic spacings of the Ag substrate. (**c**) Corresponding STM image showing macroscopic preferential uniaxial alignment of the zigzag chains parallel to the step direction (STM parameters: size = 50 x 50 nm$^2$, V = -393.7 mV; I = 234 pA). Panels (**d**) and (**e**) show respectively the LEED patterns of the pristine silver substrate before and after forming zigzag chains at the (111) region. At this position, a hexagonal pattern around the (0,0) spot is observed which evidences that the chains primarily follow three main directions. This is accordingly observed in the corresponding STM image in (**f**), where 120° rotated patches co-exist as highlighted in green, yellow and purple. (STM parameters: size = 35 x 35 nm$^2$, V = 425.2 mV; I = 254 pA).

## II. ARPES band structure and photoemission intensity simulations using the Electron Plane Wave Expansion (EPWE) method

The ARPES band structure for the zigzag chains presented in Fig.1 of the main manuscript, corresponds to the second derivative treatment of the photoemission intensity. Such data treatment is used to enhance weak photoemission features. For completeness, the raw data is included in Fig. S3.

To capture the photoemission intensity modulations observed in ARPES, we perform photoemission intensity simulations using the EPWE method[3]. These simulations (excluding the substrate and Br atoms presence) match the experimental data in Fig. S4 and support the argument of the scarce effects that the underlying Ag substrate and Br atoms have on the electronic properties of the zigzag chains.

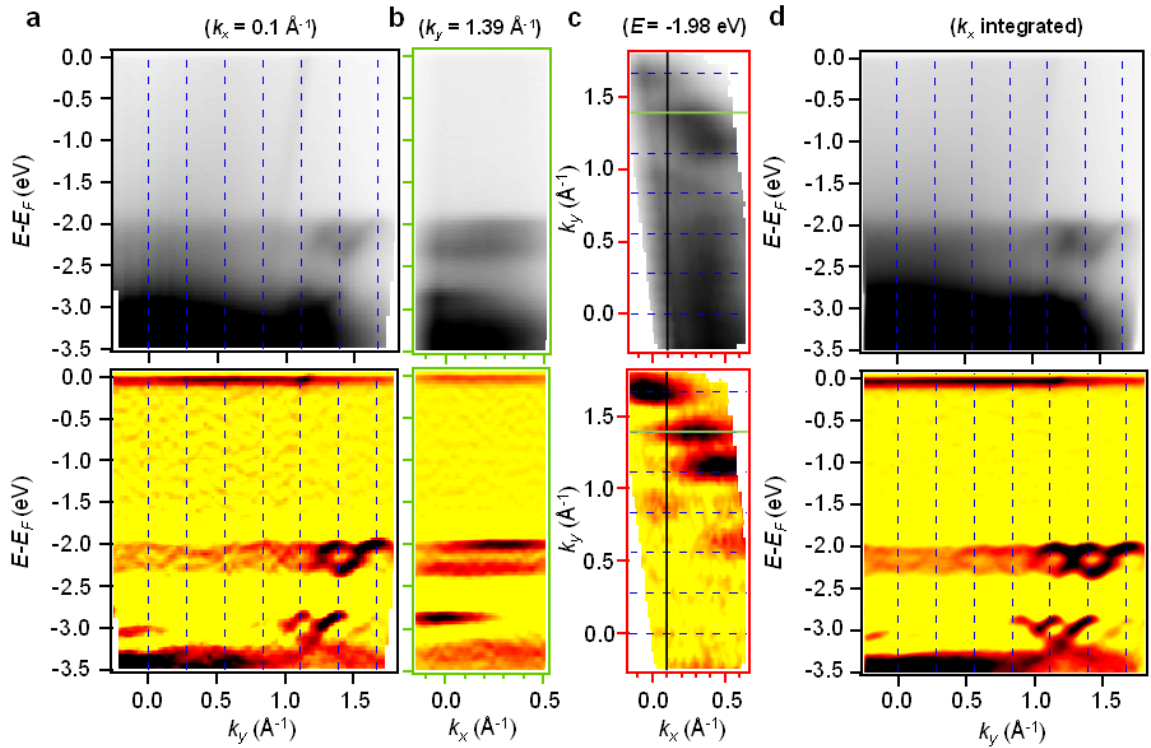

***Figure S3 | Raw data vs 2nd derivative treatment comparison for the electronic band structure of zigzag chains at the vicinal plane ~3.6° off the (111) region.*** *The ARPES spectral intensity raw data is shown on the top row and the 2nd derivative in the bottom row.* ***(a)*** *ARPES band maps (E vs $k_y$) for $k_x$ = 0.1 Å$^{-1}$.* ***(b)*** *Band structure perpendicular to the chains average axis (E vs $k_x$ with $k_y$ = 1.39 Å$^{-1}$).* ***(c)*** *Isoenergetic cut ($k_x$ vs $k_y$) at the top of the VB (E = -1.98 eV). The photoemission intensity presented in (a) has been obtained by following the direction highlighted by the vertical black line. The band structure perpendicular to the main zigzag chain axis shown in (b) is shown as a horizontal green line.* ***(d)*** *The $k_x$ integrated ARPES spectral intensity is shown to better capture all the replicating bands, which follow the periodicity of the zigzag chain unit cell (2π/L) marked with vertical dashed blue lines.*

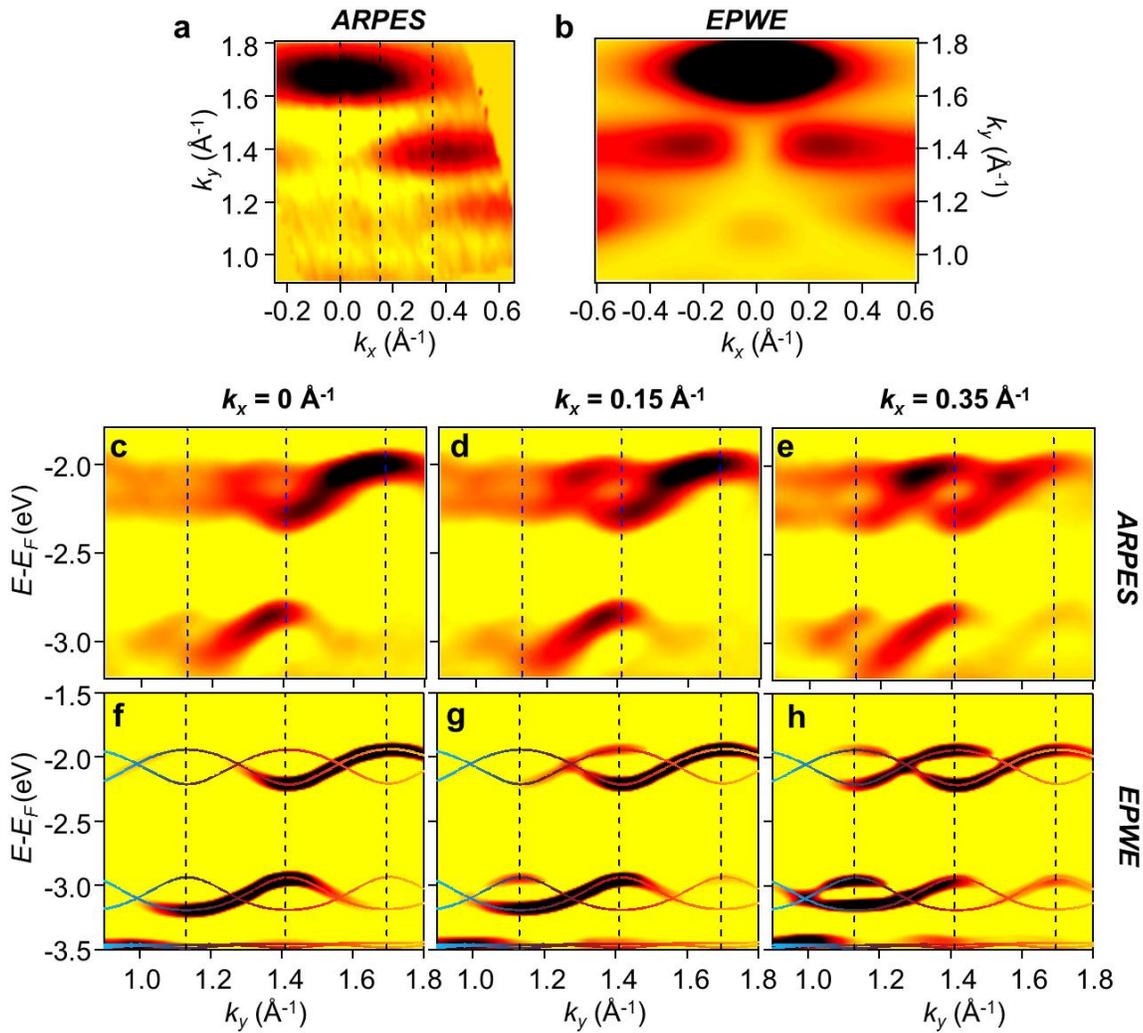

***Figure S4 | Experimental ARPES data comparison to EPWE photoemission intensity simulations.*** *The top row (**a,b**) shows the 2<sup>nd</sup> derivative of the ARPES and EPWE isoenergetic cuts ($k_x$ vs $k_y$) performed at the top of the VB (E=-1.98 eV in ARPES). The middle row (**c-e**) corresponds to the 2<sup>nd</sup> derivative of the experimental ARPES spectral intensities (E vs $k_y$) at different $k_x$ values (indicated as dashed lines in (a)). The bottom row (**f-h**) shows the equivalent photoemission intensity simulations performed with the EPWE code. Their match is exceptional, supporting the idea of weak electronic interactions between the zigzag chains with the substrate and surrounding Br atoms.*

### III. Br effect on the electronic structure of zigzag chains studied by STM/STS

After the Ullmann coupling reaction takes place and the precursors covalently bond, zigzag chains condense into islands. The split Br atoms accumulate between the chains (Fig.1 in the main manuscript and Fig. S5). These observations have been already reported for zigzag chains grown on Cu(111), hyperbenzene and honeycombene close rings on Ag(111) and other covalent nanostructures such as PPP chains[4,5,6,7]. The Br atoms remain on the Ag(111) up to ~600 K and can likely stabilize the long-range ordered zigzag structures observed in the present work. According to Merino and co-workers[7] the desorption of Br atoms intercalated in between the chains produces a lowering of the work function and a shift of the electronic states to higher binding energies without affecting their effective mass. We obtain similar results when we compare dI/dV spectra of tip manipulated isolated chains with island condensed ones (Fig. S5). In particular, we observe a rigid shift of the frontier orbitals to higher energies by 200±50 mV, implying that the $N$=4 zigzag chain energy gap (~3.7 eV) is unchanged. This evidences the limited effect of the halogen atoms on the electronic properties of the polymers.

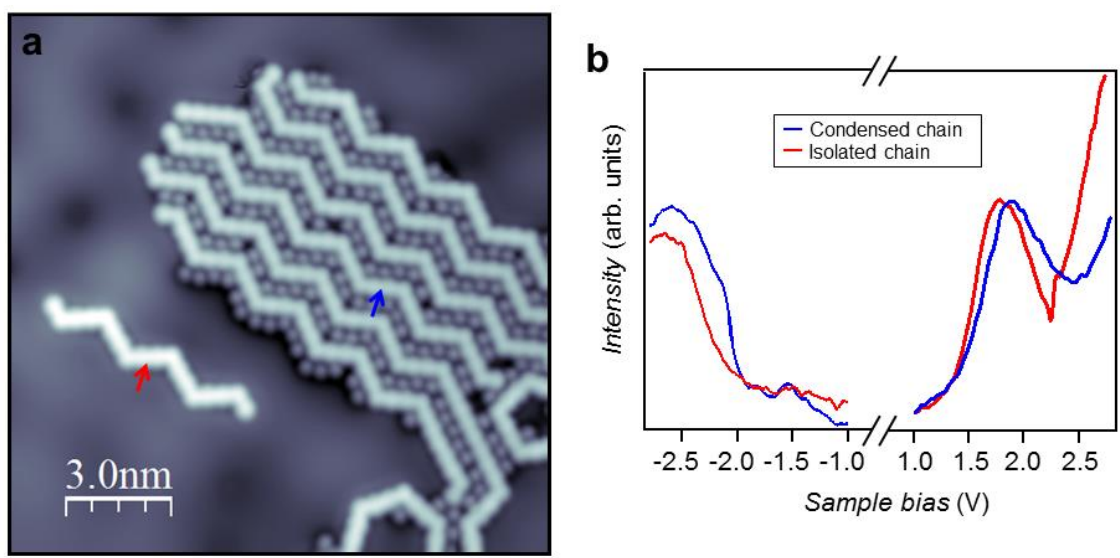

*Figure S5 │ Effect of Br adatoms on the electronic structure of isolated and condensed chains. (a)* STM image of a Br stabilized zigzag chain island (condensed chains) and an isolated, Br-free zigzag chain obtained via tip manipulation (Imaging parameters: 50 mV, 100 pA; frame: 15 x 15 nm²). *(b)* Corresponding constant-height dI/dV spectra for the VB and CB at selected positions. The presence of Br rigidly shifts the VB and CB onsets by 200 ± 50 mV to higher energies (STS parameters: Bias voltage modulation of 10 mV$_{rms}$ at 341 Hz. Close-feedback parameters: -350 mV, 150 pA and 1200 mV, 100 pA for the negative and positive resonances regions, respectively).

### IV. ARPES band structure comparison of zigzag chains vs poly-(para-phenylene) (PPP) chains.

The occupied experimental band structure of the zigzag chains is compared to the straight PPP chains in Figure S6. The PPP polymer film has likewise been grown using a vicinal Ag substrate with (554) orientation and featuring an average terrace width of 2.4 nm terminated by straight steps rotated 30° from the fully kinked curved Ag(111) (see Figure S1). Indeed, the step direction of Ag(554) follows the [1-10] direction and the terraces are slightly smaller than the one used for the zigzag, but can still host in average 3-4 parallel chains. The difference in the electronic nature between both chain types is evident from the dataset and closely follows the DFT calculations reported in Fig. 2 of the main text.

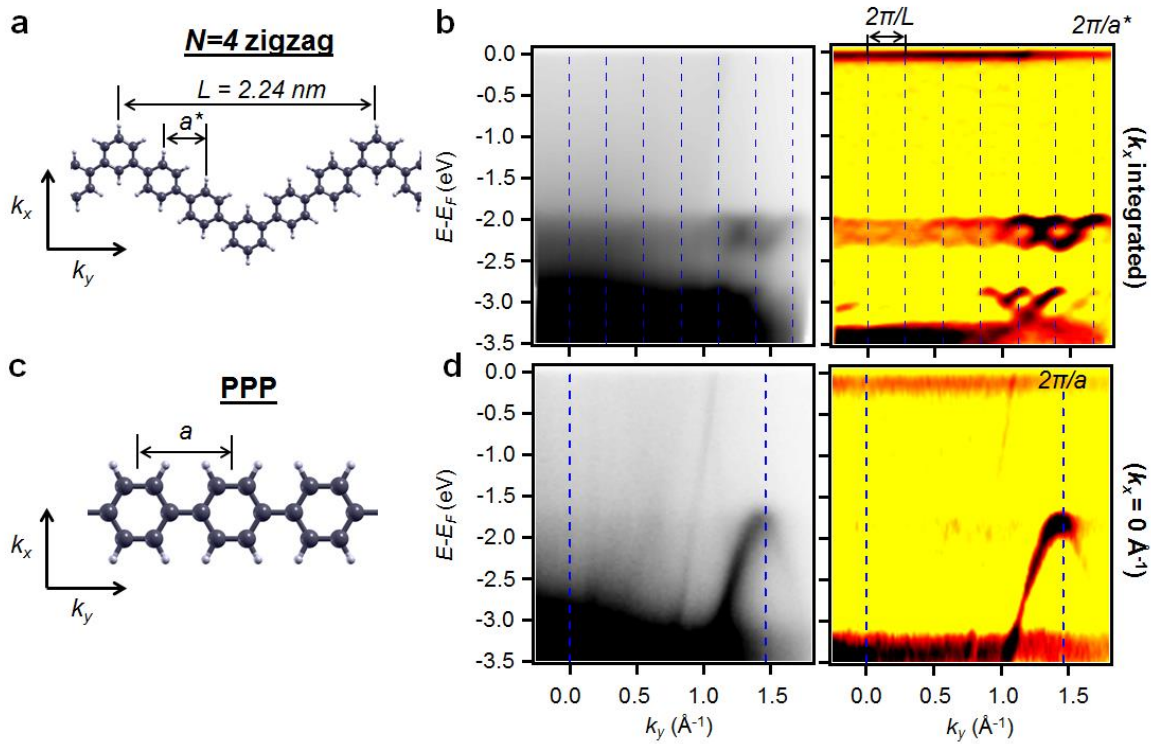

**Figure S6 │ N=4 zigzag chain ARPES band structure comparison with that of PPP chains.** *Schematic representations of the N=4 zigzag chains (a) and the poly-(para-phenylene) (PPP) chains (c). (b) and (d) show in the same row their respective raw (left) and second derivative (right) of the $k_x$ integrated ARPES spectral intensity (E vs $k_y$) along the chain axis. These plots follow the periodicity of the chain unit cells ($2\pi/L$) marked with vertical dashed blue lines. For the PPP polymer film the substrate used is a Ag(544) vicinal crystal (see text for details).*

## V. Band structure variations with straight segment's length of the zigzag chains from DFT calculations

To further investigate the evolution of the frontier orbital bandgap and electronic structure with the straight segment's length, we perform additional DFT calculations. As indicated in the main manuscript and depicted in Figure S7, zigzag chains with $N=4$ bear a closer resemblance to poly-(*meta*-phenylene) (PMP) rather than to the straight, highly dispersive polymer poly-(*para*-phenylene) (PPP) counterpart. Moreover, a gradual reduction of the electronic bandgap of the frontier orbitals is already observed (PMP > zigzag $N=4$ > zigzag $N=7$ > PPP) as the straight segments are enlarged. Such an evolution is clearly shown in Fig. S8 for a varying number of phenyl rings from 4 up to 11 in a planar, free-standing configuration. In agreement with previous work[9], we find that the band gap shrinks linearly as $1/N$ as the size of the straight segment is increased. This also agrees with our experimental results of Fig. 4c of the main manuscript, which are represented on the right axis of the graph. This evidences that while the electronic band dispersion is governed by the periodically spaced *meta*-junctions, the bandgap varies with the size of the straight segments.

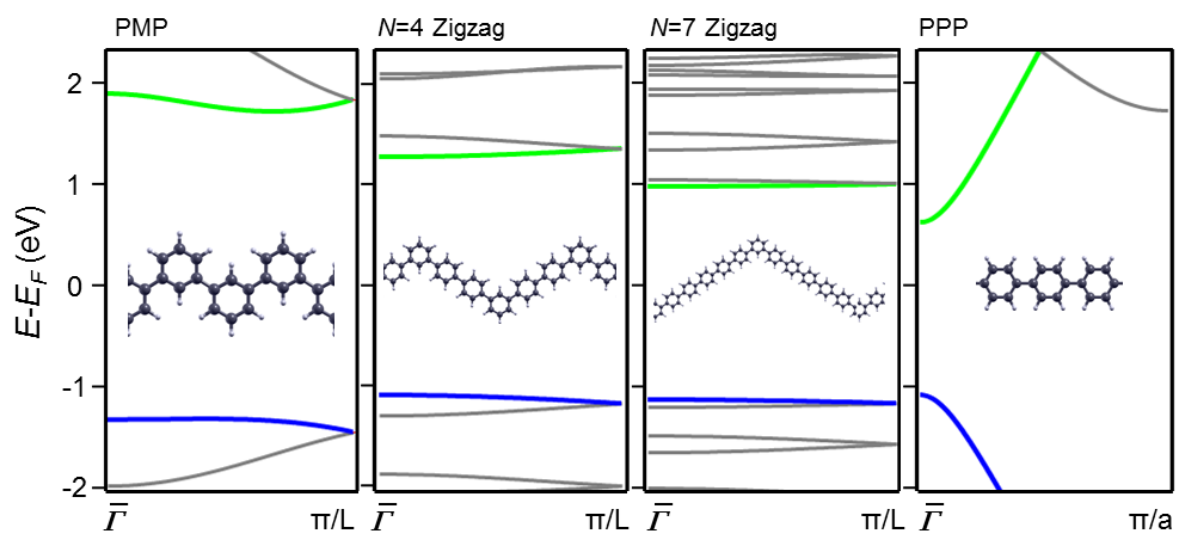

***Figure S7 | Band structure comparison between PMP, zigzag and PPP chains, as obtained by DFT calculations.*** *The highly dispersive band structure of PPP chains (at the right) clearly show a different behavior to the other three cases. These plots show a progressive reduction of the electronic bandgap as the meta-junctions (elbows) are separated further. We attribute this to the cross-conjugated nature of the polymers hosting periodically spaced meta-junctions.*

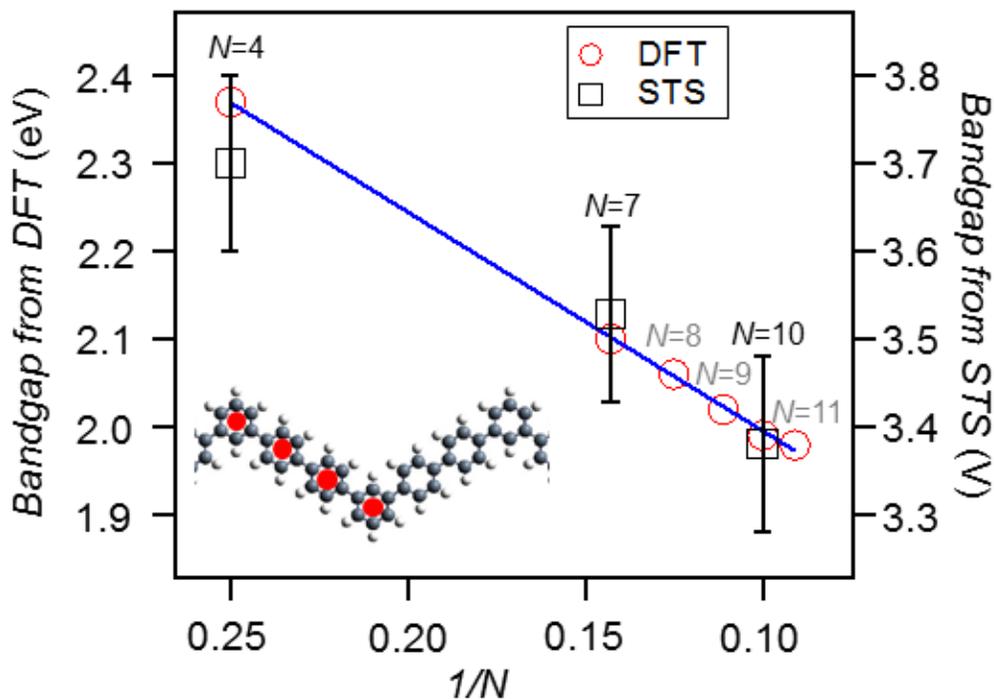

***Figure S8 | Evolution of the zigzag chain frontier orbital bandgap with increasing number of phenyl rings at the straight segments.*** *Both DFT (theory) and STS (experimental) show a clear 1/N dependency of the bandgap with increasing number of phenyl rings inside the straight segments.*

## VI. Phenyl twisting considerations of zigzag chains by STM/STS and DFT

Finally, it is worth considering the electronic effects that phenyl-phenyl twisting has on the on-surface zigzag chains. In order to reveal whether such morphological effects are present, we acquired STM images with bond resolution on condensed zigzag chains (surrounded by Br atoms) that are compared with tip manipulated isolated chains (Fig. S9). Both types of 1D structures show some very weak intensity variations at the phenyl rings, which suggest that there are slight twists in the configurations. This would agree with the observations for poly-(*para*-phenylene) PPP chains where a phenyl twisting of 20 ± 5 degrees was observed with respect to the planar configuration[8]. Indeed, the surrounding morphological conditions are determinant, since the twisting is different when the zigzag chains are condensed into islands (asymmetric arms) or are isolated (symmetric twisting of consecutive arms). Such subtle variations could be caused by the presence of Br, which could induce intermolecular interactions, or by a different matching to the underlying substrate.

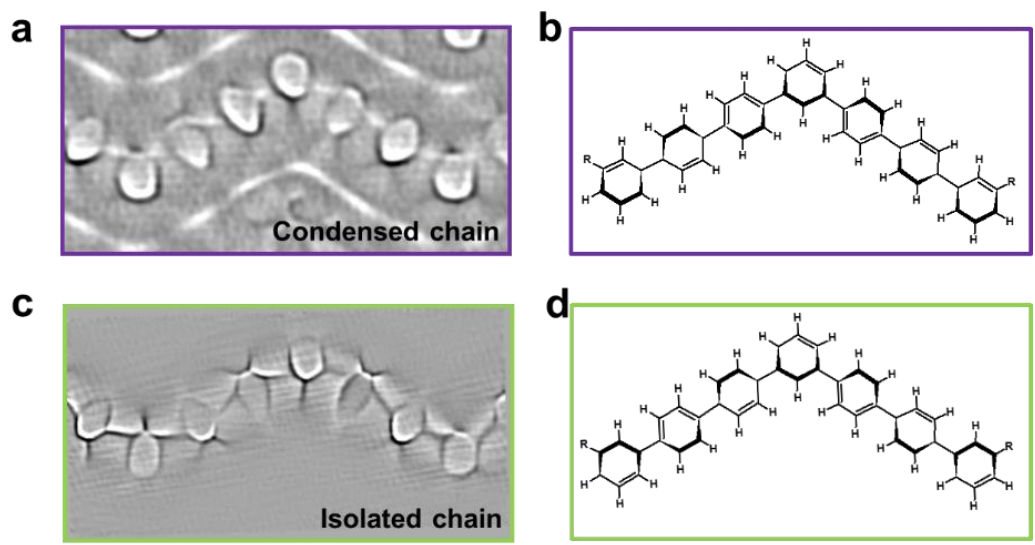

*Figure S9 | Phenyl-phenyl twisting of condensed and isolated zigzag chains.* Bond-resolution STM images (current signal, post-processed with a Laplace filter) acquired at constant height with a CO-functionalized STM-tip (frames: 3.0 x 1.5 nm$^2$, $V_{tip}$ = 2 mV) of zigzag chains condensed within an island **(a)** and an isolated chain obtained by tip manipulation **(c)**. These images suggest a slight phenyl twisting, which for the condensed chain is asymmetric with respect to the central meta-junction of both straight segments (left ascending and right descending), as shown in **(b)**, whereas it is symmetric with respect to the central meta-junction for both straight arms for the isolated chain, as shown in **(d)**.

The effect that the phenyl twisting has on the electronic structure can be clarified by DFT calculations. We focus on the *N*=4 zigzag chain and we consider three phenyl twisting configurations: a single phenyl twisting, twisting of two phenyls with the same orientation, and twisting of two phenyls with opposite orientation (see Fig. S10). As shown in Fig. S10d, in all three cases the energy minimum is found for twisting angles of approximately 20 to 35 degrees with respect to the planar configuration. However, according to the results in Figure S10e, such deviations from planarity would only induce a slight increase (by about 7%) of the frontier orbital bandgap. Upon adsorption on the substrate, the polymers are known to flatten (phenyl twist of approx. 20 degrees or less), resulting in even smaller variations of the bandgap.

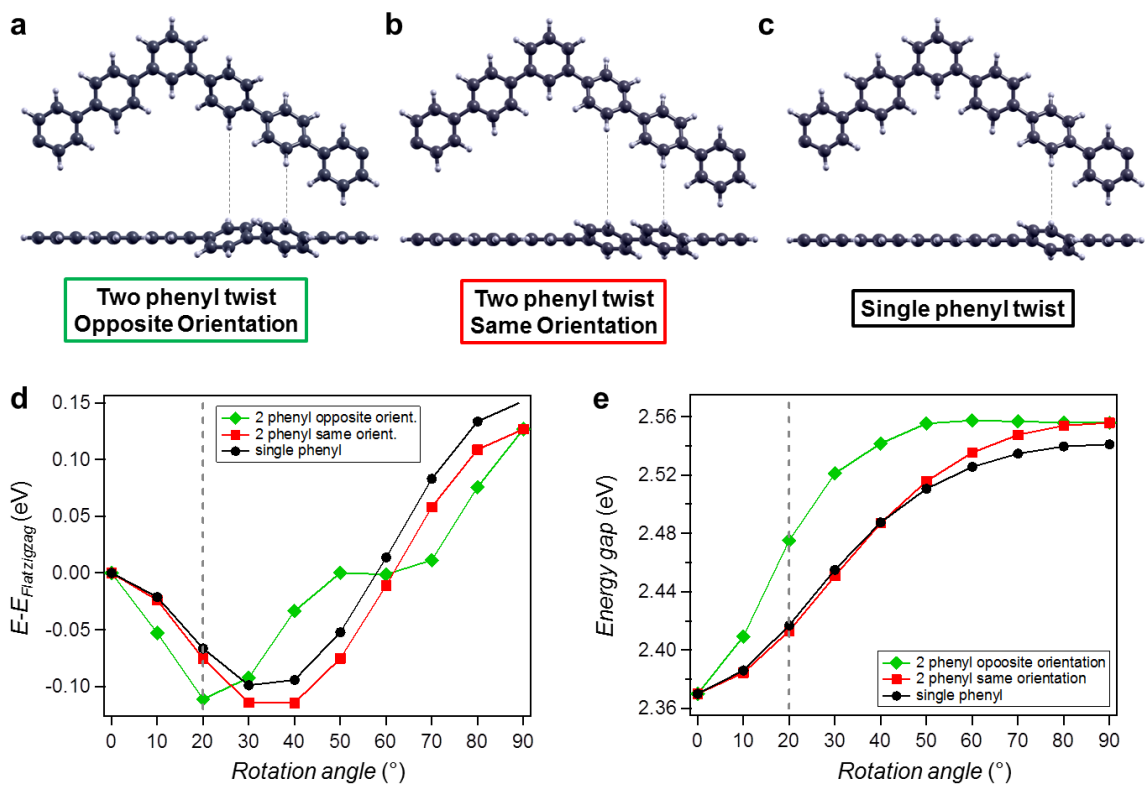

*Figure S10 │ Influence of phenyl-phenyl twisting on the stability of the zigzag chains and their electronic bandgap.* Three different phenyl twisting configurations have been considered: two-phenyl twist (opposite orientation) **(a)**, two-phenyl twist (same orientation) **(b)** and single phenyl twist **(c)**. **(d)** Energy of the three twisting configurations with respect to the fully planar configuration (zero energy). **(e)** Evolution of the electronic bandgap with respect to the phenyl rotation angle, for the three different configurations.

## References

(1) Ortega, J.E.; Vasseur, G.; Piquero-Zulaica, I.; Matencio, S.; Valbuena, M.; Rault, J.E.; Schiller, F.; Corso, M.; Mugarza, A.; Lobo-Checa, J. Structure and electronic states of vicinal Ag(111) surfaces with densely kinked steps. *New Journal of Physics* **2018**, 20, 073010.

(2) Hieulle, J.; Carbonell-Sanromá, E.; Vilas-Varela, M.; Garcia-Lekue, A.; Guitián, E.; Peña, D.; Pascual, J.I. On-Surface Route for Producing Planar Nanographenes with Azulene Moieties. *Nano Letters* **2018**, 18, 418–423.

(3) Abd El-Fattah, Z. M.; Kher-Elden, M. A.; Piquero-Zulaica, I.; García de Abajo, F. J.; Ortega, J. E. Graphene: Free electron scattering within an inverted honeycomb lattice. *arXiv:1808.06034*

(4) Koch, M.; Gille, M.; Viertel, A.; Hecht, S.; Grill, L. Substrate-controlled linking of molecular building blocks: Au(111) vs. Cu(111). *Surface Science* **2014**, 627, 70–74.

(5) Fan, Q.; Wang, C.; Han, Y.; Zhu, J.; Kuttner, J.; Hilt, G.; Gottfried, J.M. Surface-Assisted Formation, Assembly, and Dynamics of Planar Organometallic Macrocycles and Zigzag Shaped Polymer Chains with C–Cu–C Bonds. *ACS Nano* **2014**, 8, 709–718.

(6) Chen, M.; Shang, J.; Wang, Y.; Wu, K.; Kuttner, J.; Hilt, G.; Hieringer, W.; Gottfried, J.M. On-Surface Synthesis and Characterization of Honeycombene Oligophenylene Macrocycles. *ACS Nano* **2017**, 11, 134–143.

(7) Merino-Díez, N.; Lobo-Checa, J.; Nita, P.; Garcia-Lekue, A.; Basagni, A.; Vasseur, G.; Tiso, F.; Sedona, F.; Das, P.K.; Fujii, J.; Vobornik, I.; Sambi, M.; Pascual, J.I.; Ortega, J.E.; de Oteyza, D.G. Switching from Reactant to Substrate Engineering in the Selective Synthesis of Graphene Nanoribbons. *The Journal of Physical Chemistry Letters* **2018**, 9, 2510–2517.

(8) Basagni, A.; Vasseur, G.; Pignedoli, C.A.; Vilas-Varela, M.; Peña, D.; Nicolas, L.; Vitali, L.; Lobo-Checa, J.; de Oteyza, D.G.; Sedona, F.; Casarin, M.; Ortega, J.E., Sambi, M. Tunable Band Alignment with Unperturbed Carrier Mobility of On-Surface Synthesized Organic Semiconducting Wires. *ACS Nano* **2016,** 10, 2644–2651.

(9) Wang, S.; Wang, W.; Lin, N. Resolving Band-Structure Evolution and Defect-Induced States of Single Conjugated Oligomers by Scanning Tunneling Microscopy and Tight-Binding Calculations. *Physical Review Letters* **2011**, 106.